\documentclass{amsart}
\newtheorem{thm}{Theorem}

\newtheorem{lemma}{Lemma}
\newtheorem{cor}{Corollary}

\newcommand{\is}{\int_{S^{1}}}
\renewcommand{\a}{\alpha}

\newcommand{\sm}{\mathcal{S}}

\begin{document}
\title{Asymptotic expansions close to the 
singularity in Gowdy spacetimes}
\author{Hans Ringstr\"{o}m}
\address{Max-Planck-Institut f\"{u}r Gravitationsphysik, Am M\"{u}hlenberg 1,
D-14476 Golm, Germany}

\begin{abstract}
We consider Gowdy spacetimes under the assumption that the spatial
hypersurfaces are diffeomorphic to the torus. The relevant equations
are then wave map equations with the hyperbolic space as a target. 
In an article by Grubi\v{s}i\'{c} and Moncrief, a formal expansion
of solutions in the direction toward the singularity was proposed. 
Later, Kichenassamy and Rendall constructed a family of
real analytic solutions with the maximum number of free functions
and the desired asymptotics at the singularity. The condition of 
real analyticity was subsequently removed by Rendall. In an article
by the author, it was shown that one can put a condition on initial
data that leads to asymptotic expansions. In this article, we show
the following. By fixing a point in hyperbolic space, we can consider
the hyperbolic distance from this point to the solution at a given
spacetime point. If we fix a spatial point 
for the solution, it is enough to put conditions on the rate at which
the hyperbolic distance tends to infinity as time tends to the
singularity in order to conclude that there are smooth expansions in
a neighbourhood of the given spatial point.
\end{abstract}
\maketitle

\section{Introduction}

The motivation for studying the problem discussed in this article
is the desire to understand the structure of singularities in 
cosmological spacetimes. By the singularity theorems, cosmological
spacetimes typically have a singularity in the sense of causal geodesic
incompleteness. However, it seems that the methods used to obtain this
result are not so well suited to answering related questions concerning
e.g. curvature blow up. To proceed, it seems difficult to avoid
analyzing the equations in detail. After some appropriate choice of 
gauge, one is then confronted with the task of analyzing the
asymptotics of a non-linear hyperbolic equation. This is in general
not so easy. Consequently, one often imposes some symmetry condition,
and we will here consider a class of spacetimes with a two dimensional
group of symmetries. The problem one ends up with is then a system
of non-linear wave equations in $1+1$ dimensions. 

The Gowdy spacetimes were first 
introduced in \cite{gowdy} (see also \cite{chr1}), and in \cite{mon}
the fundamental questions concerning global
existence were answered. We will take the Gowdy vacuum spacetimes on
$\mathbb{R}\times T^{3}$ to be metrics of the form (\ref{eq:gowdy}). 
We do not wish to motivate this choice at length here, but refer
the reader to the above mentioned references for further details.
A brief justification is given in the introduction of \cite{jagg}.
Let
\begin{equation}\label{eq:gowdy}
 g=e^{(\tau-\lambda)/2}(-e^{-2\tau}d\tau^{2}+d\theta^2)
+e^{-\tau}[e^{P}d\sigma^2+2e^{P}Qd\sigma d\delta+
(e^{P}Q^2+e^{-P})d\delta^2]
\end{equation}
Here, $\tau\in\mathbb{R}$ and $(\theta,\sigma,\delta)$ are coordinates
on $T^{3}$. The evolution equations become
\begin{eqnarray}
P_{\tau\tau}-e^{-2\tau}P_{\theta\theta}-
e^{2P}(Q_{\tau}^2-e^{-2\tau}Q_{\theta}^2) & = & 0 \label{eq:g1}\\
Q_{\tau\tau}-e^{-2\tau}Q_{\theta\theta}
+2(P_{\tau}Q_{\tau}-e^{-2\tau}P_{\theta}Q_{\theta}) & = & 0,\label{eq:g2}
\end{eqnarray}
and the constraints
\begin{eqnarray}
\lambda_{\tau} & = & P_{\tau}^{2}+e^{-2\tau}P_{\theta}^{2}+
e^{2P}(Q_{\tau}^{2}+e^{-2\tau}Q_{\theta}^{2})\label{eq:gc1}\\
\lambda_{\theta} & = & 2(P_{\theta}P_{\tau}+e^{2P}Q_{\theta}Q_{\tau}).
\label{eq:gc2}
\end{eqnarray}
Obviously, the constraints are decoupled from the evolution equations, 
excepting the condition on $P$ and $Q$ implied by (\ref{eq:gc2}).
Thus the equations of interest are the two non-linear coupled wave
equations (\ref{eq:g1})-(\ref{eq:g2}). In the above parametrization, the
singularity corresponds to $\tau\rightarrow \infty$, and the subject
of this article is the asymptotics of solutions to 
(\ref{eq:g1})-(\ref{eq:g2}) as $\tau\rightarrow \infty$. The equations
(\ref{eq:g1})-(\ref{eq:g2}) are wave map equations. In fact, let
\[
g_{0}=-e^{-2\tau}d\tau^{2}+d\theta^{2}+e^{-2\tau}d\chi^{2}
\]
be a metric on $\mathbb{R}\times T^{2}$ and let
\begin{equation}\label{eq:grdef}
g_{R}=d P^{2}+e^{2P}d Q^{2}
\end{equation}
be a metric on $\mathbb{R}^{2}$. Then (\ref{eq:g1})-(\ref{eq:g2})
are the wave map equations for a map from $(\mathbb{R}\times T^{2},
g_{0})$ to $(\mathbb{R}^{2},g_{R})$ which is independent of the 
$\chi$-coordinate. Note that $(\mathbb{R}^{2},g_{R})$ is isometric
to the upper half plane $H=\{ (x,y)\in \mathbb{R}^{2}:y>0\}$ with metric
\begin{equation}\label{eq:ghdef}
g_{H}=\frac{d x^{2}+d y^{2}}{y^{2}}
\end{equation}
under the map 
\begin{equation}\label{eq:phirdef}
\phi_{RH}(Q,P)=(Q,e^{-P}).
\end{equation}
Thus the target space is hyperbolic space. For the rest of the article,
we will be concerned with the above mentioned wave map equations
and not consider the consequences for the resulting spacetimes.
The implications for the spacetime geometry can be found elsewhere,
see e.g. \cite{jagg}. 

The idea of finding expansions for solutions close to the 
singularity started with the article \cite{grub} by Grubi\v{s}i\'{c}
and Moncrief. In our setting, the natural expansions are 
\begin{eqnarray}
P(\tau,\theta) & = & v(\theta)\tau+\phi(\theta)+e^{-\epsilon\tau}
u(\theta,\tau)\label{eq:as1}\\
Q(\tau,\theta) & = & q(\theta)+e^{-2v(\theta)\tau}[\psi(\theta)+
w(\tau,\theta)]\label{eq:as2}
\end{eqnarray}
where $\epsilon>0$ and $w,u\rightarrow 0$ as $\tau\rightarrow \infty$ and
$0<v(\theta)<1$. This should be compared with (5) and (6) of 
\cite{kar1}, where $-Z=P$, $X=Q$ and $t=e^{-\tau}$. 
A heuristic argument motivating the condition on 
the velocity can be found in \cite{bag}. In the non-generic case
$Q=0$, one can prove that (\ref{eq:as1}) holds without the condition
on $v$. This special case is called polarized Gowdy and has been
studied in \cite{jam}, which also considers the other topologies for 
Gowdy spacetimes. The numerical simulations indicate that there
are exceptions to (\ref{eq:as1})-(\ref{eq:as2}), called spikes, at 
which the kinetic energy of the wave map makes a jump. We refer the
reader to \cite{bag} for numerical and to \cite{raw}
for analytical results concerning spikes. 

By the so called Fuchsian techniques one can construct a large family
of solutions with the asymptotic behaviour
(\ref{eq:as1})-(\ref{eq:as2}). In fact, given functions
$v,\phi,q$ and $\psi$ from $S^{1}$ to $\mathbb{R}$ of a suitable
degree of smoothness and subject to the condition $0<v<1$, one can
construct solutions to (\ref{eq:g1})-(\ref{eq:g2}) with asymptotics
of the form (\ref{eq:as1})-(\ref{eq:as2}). The proof of this in the
real analytic case can be found in \cite{kar1} and \cite{r} covers
the smooth case. One nice feature of this construction is the fact
that one gets to specify four functions freely, just as as if though
one were specifying initial data for (\ref{eq:g1})-(\ref{eq:g2}).

Finally, in \cite{jagg} a condition on the initial data leading to the 
desired asymptotics is given. Note also the related work
in \cite{chr2}. In \cite{jagg}, conditions on the $H^{2}$
norm of the first derivatives are imposed. The reason is the
following. It turns out that the most important quantity to control
is the velocity $v$ appearing in (\ref{eq:as1}). In other words, one
wants to control what $P_{\tau}$ converges to. In \cite{jagg}, this
control was achieved by integrating (\ref{eq:g1}).
In order to reach any conclusion, one then has to control
$e^{-2\tau}P_{\theta\theta}$ in the sup norm, which is the 
reason for the condition on the $H^{2}$ norm of the first
derivatives. Looking at the expansion (\ref{eq:as1}) naively, another
possibility presents itself. One expects $P_{\tau}$ to converge to 
$v$ rapidly, so that $P_{\tau}-P/\tau$ should be of the order of 
magnitude $O(\tau^{-1})$. Assume for a moment that we can prove that 
we have this estimate in the sup norm. Then one easily sees that 
$\partial_{\tau}(P/\tau)=O(\tau^{-2})$. Consequently, $P/\tau$
converges to a continuous function. Since $P_{\tau}-P/\tau$ converges
to zero, we then get the convergence of $P_{\tau}$ to a continuous 
function. What we need is consequently to prove that $P_{\tau}-P/\tau$
tends to zero as $\tau^{-1}$ under suitable circumstances. One 
quantity suited to achieve this task is
\begin{equation}\label{eq:fdef}
 F(\tau)=\frac{1}{2}\sup_{\theta\in S^{1}}
[(P_{\tau}-\frac{1}{\tau}P+e^{-\tau}
P_{\theta})^{2}
+e^{2P}(Q_{\tau}+e^{-\tau}Q_{\theta})^{2}]
\end{equation}
\[
+\frac{1}{2}\sup_{\theta\in S^{1}}
[(P_{\tau}-\frac{1}{\tau}P-e^{-\tau}
P_{\theta})^{2}+
e^{2P}(Q_{\tau}-e^{-\tau}Q_{\theta})^{2}].
\]
This object may seem unnatural at first, however it appears 
naturally in the arguments. It turns out that if 
$1\leq P\leq \tau-1$ in an interval $[\tau_{1},\tau_{2}]$, then
$F(\tau)\leq F(\tau_{1})(\tau_{1}/\tau)^{2}$ for $\tau\in [\tau_{1},
\tau_{2}]$, see Lemma \ref{lemma:decay}. This
estimate is optimal for solutions with asymptotics of the form
(\ref{eq:as1})-(\ref{eq:as2}) if we assume that $0<v<1$ and that
$\phi$ is not identically zero.

\begin{thm}\label{thm:expansions}
Consider a solution to (\ref{eq:g1})-(\ref{eq:g2}). Assume that
$\tau_{0}\geq 2$,
\begin{equation}\label{eq:pb}
1\leq P(\tau_{0},\theta)\leq \tau_{0}-1,\ \ \
\gamma\leq\frac{1}{\tau_{0}}P(\tau_{0},\theta)\leq 1-\gamma
\end{equation}
for $\theta\in S^{1}$ and some $\gamma>0$, and that 
$F(\tau_{0})\leq (\gamma-\alpha)^{2}$ for some $\a>0$, $\a<\gamma$.
Then there are $v,\phi ,q,r\in 
C^{\infty}(S^{1},\mathbb{R})$, and for all non-negative integers $k$, 
polynomials $\Xi_{i,k}$, $i=1,...,4$ in $\tau$, where
\begin{equation}\label{eq:vcon}
0<\a\leq v\leq 1-\a<1
\end{equation}
on $S^{1}$ such that
\begin{equation}\label{eq:velcon}
\| P_{\tau}- v\|_{C^{k}(S^{1},\mathbb{R})}\leq
\Xi_{1,k}\exp[-\a\tau],
\end{equation}
\begin{equation}\label{eq:pconv}
\|P-p\|_{C^{k}(S^{1},\mathbb{R})}\leq
\Xi_{2,k}\exp[-\a\tau],
\end{equation}
\begin{equation}\label{eq:xi5}
\|e^{2p}Q_{\tau}-r\|_{C^{k}(S^{1},\mathbb{R})}\leq 
\Xi_{3,k}\exp[-\a\tau],
\end{equation}
and
\begin{equation}\label{eq:xi6}
\| e^{2p}(Q-q)+\frac{r}{2v}\|_{C^{k}(S^{1},\mathbb{R})}\leq
\Xi_{4,k}\exp[-\a\tau]
\end{equation}
for all $\tau\in [\tau_{0},\infty)$, where
$p=v\cdot\tau+\phi $.
\end{thm}
\textit{Remark}. The result obtained is the same as 
in \cite{jagg}, but the conditions are weaker. We wish to emphasize
that the proof does not rely on \cite{jagg} 
in any essential way. We only need two pages of estimates from that
paper in order to complete the argument. Thus, the rather
intricate arguments presented in \cite{jagg} are avoided.

The proof is to be found at the end of Section \ref{section:est}.
Under the conditions of the theorem, we thus obtain the 
expansions (\ref{eq:as1})-(\ref{eq:as2}). Note the condition
(\ref{eq:pb}) which requires that $P\geq 1$. This is a rather
unnatural condition which one would like to get rid of. It turns
out that the reason one has to impose this condition is due to 
a bad choice of representative of hyperbolic space. Consider the
asymptotic behaviour (\ref{eq:as1})-(\ref{eq:as2}). Under the 
map (\ref{eq:phirdef}), one gets the following picture in the 
upper half plane. The $x$-coordinate converges for every fixed 
$\theta$ and the $y$-coordinate tends to zero. In other words, for 
a fixed $\theta$, the solution goes to the boundary along a geodesic,
at least approximately. Let us consider the same picture in the disc
model. Let $D$ be the open unit disc in the complex plane and let
\begin{equation}\label{eq:gddef}
g_{D}=\frac{4(d x^{2}+d y^{2})}{(1-x^{2}-y^{2})^{2}}
\end{equation}
be a metric on it. In complex notation, we have the isometry
\begin{equation}\label{eq:phihdef}
\phi_{HD}(z)=\frac{z-i}{z+i}
\end{equation}
from the upper half plane to the disc model. Note that the inverse
is given by $\phi_{HD}^{-1}(w)=i (1+w)/(1-w)$. Say now that we have
a solution tending to the boundary in the disc model. For some
spatial point, the solution tends to $1$, and for neighbouring 
points, the solution goes to the boundary a little bit below and 
a little bit above in the complex plane. If one translates this 
to the $PQ$-variables, one gets a very violent behaviour which is 
completely unrelated to the geometry of the wave map problem.
For this reason, we 
consider the equations in the disc model. We have
\begin{equation}\label{eq:edisc}
 \partial_{\tau}\left(\frac{z_{\tau}}{(1-|z|^{2})^{2}}\right)
-e^{-2\tau}\partial_{\theta}\left(\frac{z_{\theta}}{(1-|z|^{2})^{2}}\right)
=\frac{2z}{(1-|z|^{2})^{3}}[|z_{\tau}|^{2}-e^{-2\tau}|z_{\theta}|^{2}].
\end{equation}
We will use the notation
\begin{equation}\label{eq:dn}
|z_{\tau}|_{D}^{2}=\frac{4}{(1-|z|^{2})^{2}}|z_{\tau}|^{2}
\end{equation}
and similarly for $z_{\theta}$, $z_{\tau}+e^{-\tau}z_{\theta}$ etc.
Let 
\begin{equation}\label{eq:rhodef}
\rho=\ln\frac{1+|z|}{1-|z|}.
\end{equation}
This is the distance from the origin to the point $z$ with respect to
the hyperbolic metric. Note that $\rho^{2}$ and $\rho/|z|$ are smooth
functions on the open unit disc. 

\begin{thm}\label{thm:td}
Consider a solution $z$ to (\ref{eq:edisc}). Assume that there is an
$0<\a<1$ such that at $\tau_{0}$, 
\begin{equation}\label{eq:object}
\frac{1}{2}\sup_{\theta}|z_{\tau}+e^{-\tau}z_{\theta}|_{D}^{2}+
\frac{1}{2}\sup_{\theta}|z_{\tau}-e^{-\tau}z_{\theta}|_{D}^{2}\leq
(1-\a)^{2}.
\end{equation}
Then there is a $v\in C^{0}(S^{1},\mathbb{R}^{2})$
with $|v|\leq 1-\a$ and a $C$ and a $T$ such that for $\tau\geq T$, 
\[
 \|\frac{1}{\tau}\frac{z(\tau,\cdot)}{|z(\tau,\cdot)|}
\rho(\tau,\cdot)-v\|_{C^{0}(S^{1},\mathbb{R}^{2})}+
\|\frac{2z_{\tau}(\tau,\cdot)}{1-|z(\tau,\cdot)|^{2}}-
v\|_{C^{0}(S^{1},\mathbb{R}^{2})}
\]
\[
+e^{-\tau}\|\frac{2z_{\theta}(\tau,\cdot)}{1-|z(\tau,\cdot)|^{2}}
\|_{C^{0}(S^{1},\mathbb{R}^{2})}\leq C\tau^{-1}.
\]
\end{thm}
\textit{Remark}. Note that if $v(\theta_{0})=0$, then the above 
estimate implies that $z(\tau,\theta_{0})$ remains bounded as 
$\tau\rightarrow\infty$. In the proof, we show that the left hand
side of (\ref{eq:object}) decreases as $\tau$ increases. 

The proof is to be found at the end of Section \ref{section:dm}.
As noted in Lemma \ref{lemma:rhobo}, one gets the same conclusions 
if one replaces the condition (\ref{eq:object}) with the condition
that there is a $T$ such that $\rho(\tau,\theta)\leq\tau-2$ for 
$\tau\geq T$. Of course, one would like to have smooth expansions. 
Note that this can be done in all generality in the polarized case,
see \cite{jam}. We are not able to prove that there are smooth
expansions if $|v|=0$, but all other cases can be dealt with 
in the following sense. 

\begin{thm}\label{thm:local}
Consider a solution to (\ref{eq:edisc}).
Let $\theta_{0}\in S^{1}$ be a fixed angle, and assume that there is 
a $T$ and a $0<\gamma<1$ such that 
\[
\frac{\rho(\tau,\theta_{0})}{\tau}\leq 1-\gamma
\]
for all $\tau\geq T$. Assume furthermore that there is a 
sequence $\tau_{k}\rightarrow \infty$ such that 
\[
\frac{\rho(\tau_{k},\theta_{0})}{\tau_{k}}\geq \gamma
\]
Then there is an $\eta>0$ and an
isometry $\phi$ from $(D,g_{D})$ to $(\mathbb{R}^{2},g_{R})$
such that if $(Q,P)=\phi\circ z$, then the conclusions of 
Theorem \ref{thm:expansions} hold if we replace $S^{1}$ with
$I_{\eta}$, where $I_{\eta}=(\theta_{0}-\eta,\theta_{0}+\eta)$.
\end{thm}
The theorem follows from Lemma \ref{lemma:loco} and Lemma
\ref{lemma:rholbo}. 
If we skip the condition that $\rho(\tau_{k},\theta_{0})/\tau_{k}
\geq\gamma$, we still get conclusions similar to those of Theorem
\ref{thm:td}, cf. Lemma \ref{lemma:rholbo}. 

When working on this article, the author was made aware of the fact 
that similar results had been obtained by Chae and Chru\'{s}ciel.
Since their methods are quite different from ours, it is however our
hope that the reader will find both articles interesting.

\section{Estimates}\label{section:est}

The purpose of this section is to prove Theorem \ref{thm:expansions}.
We begin by describing the decay properties of the function $F$ 
defined in (\ref{eq:fdef}). 

\begin{lemma}\label{lemma:decay}
Consider a solution to (\ref{eq:g1})-(\ref{eq:g2}). Assume that
$\tau_{2}\geq\tau_{1}\geq 2$ and that 
\[
1\leq P(\tau,\theta)\leq \tau-1
\]
for $(\tau,\theta)\in [\tau_{1},\tau_{2}]\times S^{1}$. Then if
$F$ is defined in (\ref{eq:fdef}), 
\[
F(\tau)\leq F(\tau_{1})\left(
\frac{\tau_{1}}{\tau}\right)^{2}
\]
for all $\tau\in [\tau_{1},\tau_{2}]$. 
\end{lemma}
\textit{Proof}. Let
\[
\mathcal{A}_{1}=\frac{1}{2}e^{\tau}(P_{\tau}-\frac{1}{\tau}P+e^{-\tau}
P_{\theta})^{2},\ \ \
\mathcal{A}_{2}=\frac{1}{2}e^{\tau}e^{2P}(Q_{\tau}+e^{-\tau}Q_{\theta})^{2},
\]
\[
\mathcal{B}_{1}=\frac{1}{2}e^{\tau}(P_{\tau}-\frac{1}{\tau}P-e^{-\tau}
P_{\theta})^{2},\ \ \
\mathcal{B}_{2}=\frac{1}{2}e^{\tau}e^{2P}(Q_{\tau}-e^{-\tau}Q_{\theta})^{2}
\]
and
\begin{equation}\label{eq:abdef}
\mathcal{A}=\mathcal{A}_{1}+\mathcal{A}_{2}, \ \ \
\mathcal{B}=\mathcal{B}_{1}+\mathcal{B}_{2}.
\end{equation}
One can compute that 
\[
 (\partial_{\tau}-e^{-\tau}\partial_{\theta})\mathcal{A}_{1}=
(\frac{1}{2}-\frac{1}{\tau})e^{\tau}
[(P_{\tau}-\frac{1}{\tau}P)^{2}+e^{-2\tau}P_{\theta}^{2}]
-(1-\frac{2}{\tau})e^{-\tau}P_{\theta}^{2}
\]
\[
+e^{2P+\tau}(Q_{\tau}^{2}-
e^{-2\tau}Q_{\theta}^{2})(P_{\tau}-\frac{1}{\tau}P
+e^{-\tau}P_{\theta})
\]
and that 
\begin{equation}\label{eq:a2}
 (\partial_{\tau}-e^{-\tau}\partial_{\theta})\mathcal{A}_{2}=
\frac{1}{2}e^{2P+\tau}[Q_{\tau}^{2}+e^{-2\tau}Q_{\theta}^{2}]
\end{equation}
\[
-e^{2P+\tau}(P_{\tau}+e^{-\tau}P_{\theta})(Q_{\tau}^{2}-
e^{-2\tau}Q_{\theta}^{2})
-e^{2P-\tau}Q_{\theta}^{2}.
\]
Thus
\[
 (\partial_{\tau}-e^{-\tau}\partial_{\theta})\mathcal{A}=
 (\frac{1}{2}-\frac{1}{\tau})e^{\tau}
[(P_{\tau}-\frac{1}{\tau}P)^{2}+e^{-2\tau}P_{\theta}^{2}]
-(1-\frac{2}{\tau})e^{-\tau}P_{\theta}^{2}
\]
\[
-\frac{P}{\tau}e^{2P+\tau}(Q_{\tau}^{2}
-e^{-2\tau}Q_{\theta}^{2})
+\frac{1}{2}e^{2P+\tau}[Q_{\tau}^{2}+e^{-2\tau}Q_{\theta}^{2}]
-e^{2P-\tau}Q_{\theta}^{2}.
\]
Assuming $1\leq P\leq \tau-1$ and $\tau\geq 2$, we thus get
\[
(\partial_{\tau}-e^{-\tau}\partial_{\theta})\mathcal{A}\leq
 (\frac{1}{2}-\frac{1}{\tau})(\mathcal{A}+\mathcal{B}).
\]
Similarly, if $1\leq P\leq \tau-1$ and $\tau\geq 2$,
\[
(\partial_{\tau}+e^{-\tau}\partial_{\theta})\mathcal{B}\leq
 (\frac{1}{2}-\frac{1}{\tau})(\mathcal{A}+\mathcal{B}).
\]
Let us write down the analogue of (\ref{eq:a2}) for future 
reference.
\begin{equation}\label{eq:b2}
 (\partial_{\tau}+e^{-\tau}\partial_{\theta})\mathcal{B}_{2}=
\frac{1}{2}e^{2P+\tau}[Q_{\tau}^{2}+e^{-2\tau}Q_{\theta}^{2}]
\end{equation}
\[
-e^{2P+\tau}(P_{\tau}-e^{-\tau}P_{\theta})(Q_{\tau}^{2}-
e^{-2\tau}Q_{\theta}^{2})
-e^{2P-\tau}Q_{\theta}^{2}.
\]
Let
\[
F_{1}(u,\theta)=\mathcal{A}(u,\theta+e^{-u}),\ \ \
F_{2}(u,\theta)=\mathcal{B}(u,\theta-e^{-u}),
\]
\begin{equation}\label{eq:edef}
\hat{F}_{i}(u)=\sup_{\theta\in S^{1}}F_{i}(u,\theta),\ \ \
\hat{F}(u)=\hat{F}_{1}(u)+\hat{F}_{2}(u).
\end{equation}
We get
\[
 F_{1}(u_{2},\theta)=F_{1}(u_{1},\theta)+
\int_{u_{1}}^{u_{2}}[(\partial_{u}-e^{-u}\partial_{\theta})
\mathcal{A}](u,\theta+e^{-u})d u
\]
\[
\leq \hat{F}_{1}(u_{1})+\int_{u_{1}}^{u_{2}}
(\frac{1}{2}-\frac{1}{u})\hat{F}(u)d u,
\]
assuming $1\leq P(\tau,\theta)\leq \tau-1$ and $\tau\geq 2$ for
$(\tau,\theta)\in [u_{1},u_{2}]\times S^{1}$.
Take the sup norm of this and add it to a similar estimate for 
$\hat{F}_{2}$ in order to obtain
\[
\hat{F}(u_{2})\leq \hat{F}(u_{1})+\int_{u_{1}}^{u_{2}}
(1-\frac{2}{u})\hat{F}(u)d u.
\]
Gr\"{o}nwall's lemma implies
\[
e^{-u_{2}}\hat{F}(u_{2})\leq e^{-u_{1}}\hat{F}(u_{1})\left(
\frac{u_{1}}{u_{2}}\right)^{2}.
\]
The lemma follows. $\hfill\Box$

\begin{cor}\label{cor:continuation}
Consider a solution to (\ref{eq:g1})-(\ref{eq:g2}). Assume that
$\tau_{0}\geq 2$,
\[
1\leq P(\tau_{0},\theta)\leq \tau_{0}-1,\ \ \
\gamma\leq\frac{1}{\tau_{0}}P(\tau_{0},\theta)\leq 1-\gamma
\]
for $\theta\in S^{1}$ and some $\gamma>0$, and that 
$F(\tau_{0})\leq (\gamma-\alpha)^{2}$ for some $\a>0$, $\a<\gamma$. 
Then
\[
F(\tau)\leq F(\tau_{0})\left(
\frac{\tau_{0}}{\tau}\right)^{2}
\]
for all $\tau\geq \tau_{0}$. Furthermore, there is a function
$v\in C^{0}(S^{1},\mathbb{R})$ and a constant $0<C<\infty$ such that 
\[
\| P_{\tau}(\tau,\cdot)-v\|_{C^{0}(S^{1},\mathbb{R})}\leq
\frac{C}{\tau}
\]
for $\tau\geq \tau_{0}$ and 
$\a\leq v(\theta)\leq 1-\a$
for all $\theta\in S^{1}$.
\end{cor}
\textit{Proof}. Consider the set 
\[
\mathcal{S}=\{ \tau\geq \tau_{0}:s\in [\tau_{0},\tau]\Rightarrow
1\leq P(s,\theta)\leq s-1\ \ \forall \theta\in S^{1}\}.
\]
This set is closed and connected by definition. From the conditions of
the lemma, it is clear that it is non-empty. We wish to prove that it
is open (in the subspace topology on $[\tau_{0},\infty)$). Assume that
$\tau_{1}\in\mathcal{S}$. Let us estimate, using Lemma
\ref{lemma:decay},
\begin{equation}\label{eq:it} 
|\frac{P(\tau_{1},\theta)}{\tau_{1}}-\frac{P(\tau_{0},\theta)}{\tau_{0}}|=
|\int_{\tau_{0}}^{\tau_{1}}\frac{1}{\tau}[P_{\tau}(\tau,\theta)-
\frac{1}{\tau}P(\tau,\theta)]d\tau |\leq
\int_{\tau_{0}}^{\tau_{1}}\frac{F^{1/2}(\tau)}{\tau}d\tau
\end{equation}
\[
\leq (\gamma-\alpha)\int_{\tau_{0}}^{\tau_{1}}\frac{\tau_{0}}{\tau^{2}}
d\tau\leq (\gamma-\alpha)(1-\frac{\tau_{0}}{\tau_{1}}).
\]
Using Lemma \ref{lemma:decay} once again we conclude that
\[
|P_{\tau}(\tau_{1},\theta)-\frac{P(\tau_{1},\theta)}{\tau_{1}}|
\leq (\gamma-\a)\frac{\tau_{0}}{\tau_{1}}.
\]
Combining these two estimates, we get 
\[
| P_{\tau}(\tau_{1},\theta)-\frac{P(\tau_{0},\theta)}{\tau_{0}}|
\leq \gamma-\alpha.
\]
Since $\gamma\leq P(\tau_{0},\theta)/\tau_{0}\leq 1-\gamma$ for all
$\theta\in S^{1}$, 
$\a\leq P_{\tau}(\tau_{1},\theta)\leq 1-\a$ for all $\theta\in S^{1}$.
This implies that there is a set of the form
$[\tau_{1},\tau_{1}+\epsilon)$ with 
$\epsilon>0$ such that $1\leq P(s,\theta)\leq s-1$ for all $s$ in this 
set and $\theta \in S^{1}$. Consequently, $\mathcal{S}$ is open, so
that $\mathcal{S}=[\tau_{0},\infty)$. An estimate similar to that of 
(\ref{eq:it}) yields the conclusion that for $\tau_{2}\geq
\tau_{1}\geq \tau_{0}$
\[
\|\frac{P(\tau_{2},\cdot)}{\tau_{2}}-
\frac{P(\tau_{1},\cdot)}{\tau_{1}}\|_{C^{0}(S^{1},\mathbb{R})}
\leq \frac{C}{\tau_{1}}.
\]
Thus there is a $v\in C^{0}(S^{1},\mathbb{R})$ such that 
\[
\|\frac{P(\tau,\cdot)}{\tau}-v\|_{C^{0}(S^{1},\mathbb{R})}\leq
\frac{C}{\tau}.
\]
Letting $\tau_{1}$ tend to infinity in (\ref{eq:it}), we get the 
conclusion that $\a\leq v(\theta)\leq 1-\a$ for all $\theta\in S^{1}$.
Combining this with the fact that $F$ decays like $\tau^{-2}$, 
we get the conclusions of the corollary. $\hfill\Box$

\begin{lemma}\label{lemma:qdecay}
Consider a solution to (\ref{eq:g1})-(\ref{eq:g2}) with 
initial data satisfying the conditions in the statement of 
Corollary \ref{cor:continuation}. Then
\[
\| e^{2P}[Q_{\tau}^{2}(\tau,\cdot)+
e^{-2\tau}Q_{\theta}^{2}(\tau,\cdot)]\|_{C^{0}(S^{1},\mathbb{R})}
\leq Ce^{-\a\tau}.
\]
\end{lemma}
\textit{Proof}. Consider (\ref{eq:a2}) and (\ref{eq:b2}). By
Corollary \ref{cor:continuation} we conclude that $e^{-\tau}P_{\theta}$
converges to zero in the sup norm. Since we also know that $P_{\tau}$
converges to $v$, we conclude that for $\tau$ great enough,
\[
\frac{\a}{2}\leq P_{\tau}\pm e^{-\tau}P_{\theta}\leq 
1-\frac{\a}{2}.
\]
Combining this with (\ref{eq:a2}) and (\ref{eq:b2}) we obtain
\[
 (\partial_{\tau}-e^{-\tau}\partial_{\theta})\mathcal{A}_{2}\leq
(\frac{1}{2}-\frac{\a}{2})(\mathcal{A}_{2}+\mathcal{B}_{2}), 
\ \ \
(\partial_{\tau}+e^{-\tau}\partial_{\theta})\mathcal{B}_{2}\leq
(\frac{1}{2}-\frac{\a}{2})(\mathcal{A}_{2}+\mathcal{B}_{2}).
\]
The remaining part of the argument is similar to the 
end of the proof of Lemma \ref{lemma:decay}. $\hfill\Box$

Consider the energies
\[
\mathcal{E}_{k}=\frac{1}{2}\is
[(\partial_{\theta}^{k}\partial_{\tau}P)^{2}+
e^{-2\tau}(\partial_{\theta}^{k+1}P)^{2}]d\theta
\]
and
\[
E_{k}=\frac{1}{2}\is e^{2P}[(\partial_{\theta}^{k}\partial_{\tau}Q)^{2}+
e^{-2\tau}(\partial_{\theta}^{k+1}Q)^{2}]d\theta.
\]
We will also be interested in the energies
\begin{equation}\label{eq:hdef}
H_{k}=1+\mathcal{E}_{k}+E_{k}+\frac{1}{2}e^{-\a\tau}
\is (\partial_{\theta}^{k}P)^{2}d\theta.
\end{equation}
Note that 
\begin{equation}\label{eq:vareder}
\frac{d\mathcal{E}_{k}}{d\tau}\leq\is
\partial_{\theta}^{k}[e^{2P}(Q_{\tau}^{2}-e^{-2\tau}Q_{\theta}^{2})]
\partial_{\theta}^{k}\partial_{\tau}Pd\theta.
\end{equation}
Furthermore, since we can assume $\alpha/2\leq P_{\tau}\leq 1-\alpha/2$
for $\tau$ large enough, one can estimate
\begin{equation}\label{eq:dekdt}
\frac{d  E_{k}}{d\tau}\leq -\alpha E_{k}+
\is f_{k}\partial_{\theta}^{k}\partial_{\tau} Qd\theta
\end{equation}
for $\tau$ large enough, where 
\begin{equation}\label{eq:fkdef}
f_{k}=\partial_{\tau}(e^{2P}\partial_{\theta}^{k}\partial_{\tau}Q)
-e^{-2\tau}\partial_{\theta}(e^{2P}\partial_{\theta}^{k+1}Q).
\end{equation}
Note that $f_{0}=0$ and that 
\begin{equation}\label{eq:fkre}
f_{k+1}=\partial_{\theta}f_{k}-2P_{\theta}f_{k}
-2P_{\theta\tau}e^{2P}\partial_{\theta}^{k}\partial_{\tau}Q+
2P_{\theta\theta}e^{2P-2\tau}\partial_{\theta}^{k+1}Q.
\end{equation}
The details of the derivations of the last three equations can 
be found in the proof of Lemma 5.2 of \cite{jagg}.

\begin{lemma}\label{lemma:hob}
Consider a solution to (\ref{eq:g1})-(\ref{eq:g2}) with 
initial data satisfying the conditions in the statement of 
Corollary \ref{cor:continuation}. Then $H_{1}$ defined by 
(\ref{eq:hdef}) is bounded to the future. 
\end{lemma}
\textit{Proof}. We wish to prove this by proving an estimate of the
form
\[
\frac{d H_{1}}{d\tau}\leq Ce^{-\beta\tau}H_{1},
\]
where $\beta>0$. Let us prove this with $H_{1}$ on the left hand side
replaced by the constituents of $H_{1}$. By (\ref{eq:vareder}) we have
\[
 \frac{d\mathcal{E}_{1}}{d\tau}\leq
\is
\partial_{\theta}[e^{2P}(Q_{\tau}^{2}-e^{-2\tau}Q_{\theta}^{2})]
P_{\tau\theta}d\theta\leq
\sqrt{2}\| \partial_{\theta}[e^{2P}(Q_{\tau}^{2}-e^{-2\tau}Q_{\theta}^{2})]
\|_{L^{2}(S^{1},\mathbb{R})}\mathcal{E}_{1}^{1/2}.
\]
Let us estimate the $L^{2}$-norm appearing on the right hand side. 
There are two cases to deal with. Estimate, using Lemma
\ref{lemma:qdecay},
\[
\| P_{\theta}e^{2P}(Q_{\tau}^{2}-e^{-2\tau}Q_{\theta}^{2})
\|_{L^{2}(S^{1},\mathbb{R})}\leq
Ce^{-\a\tau}e^{\a\tau/2}H_{1}^{1/2},
\]
which is of the desired form. The other case can be estimated
\[
\| e^{2P}(Q_{\tau}Q_{\tau\theta}-e^{-2\tau}Q_{\theta}Q_{\theta\theta})
\|_{L^{2}(S^{1},\mathbb{R})}\leq
Ce^{-\a\tau/2}E_{1}^{1/2},
\]
using Lemma \ref{lemma:qdecay}. By (\ref{eq:dekdt})-(\ref{eq:fkre})
and the fact that $f_{0}=0$, we get
\[
\frac{d  E_{1}}{d\tau}\leq
\is[-2P_{\tau\theta}e^{2P}Q_{\tau}
Q_{\tau\theta}
+2e^{2P-2\tau}P_{\theta\theta}Q_{\theta}Q_{\tau\theta}]d\theta.
\]
Thus we can use Lemma \ref{lemma:qdecay} in order to obtain
\[
\frac{d  E_{1}}{d\tau}\leq Ce^{-\a\tau/2}(E_{1}+\mathcal{E}_{1}).
\]
Finally
\[
\frac{d }{d\tau}[\frac{1}{2}e^{-\a\tau}
\is P_{\theta}^{2}d\theta]=
-\frac{\a}{2}e^{-\a\tau}
\is P_{\theta}^{2}d\theta+
e^{-\a\tau}
\is P_{\theta}P_{\tau\theta}d\theta\leq 2e^{-\a\tau/2}H_{1}^{1/2}
\mathcal{E}_{1}^{1/2}.
\]
The lemma follows. $\hfill\Box$

\begin{cor}
Consider a solution to (\ref{eq:g1})-(\ref{eq:g2}) with 
initial data satisfying the conditions in the statement of 
Corollary \ref{cor:continuation}. Then
\begin{equation}\label{eq:eodec}
E_{1}(\tau)\leq C(1+\tau^{2})e^{-\a\tau}
\end{equation}
for all $\tau\geq \tau_{0}$. 
\end{cor}
\textit{Proof}. By (\ref{eq:dekdt})-(\ref{eq:fkre}) and the fact
that $f_{0}=0$, we get
\[
\frac{d  E_{1}}{d\tau}\leq
-\a E_{1}+C\exp[-\a\tau/2]E_{1}^{1/2},
\]
where we have also used Lemma \ref{lemma:qdecay} and the fact that
$H_{1}$ is bounded. This yields the conclusion of the corollary. 
$\hfill\Box$

\begin{lemma}\label{lemma:hkb}
Consider a solution to (\ref{eq:g1})-(\ref{eq:g2}) with 
initial data satisfying the conditions in the statement of 
Corollary \ref{cor:continuation}. Then for every $k$, there is a 
constant $C_{k}$ and a polynomial $\mathcal{P}_{k}$ in $\tau$
such that for $\tau\geq 0$
\[
H_{k}\leq C_{k},\ \ \
E_{k}\leq \mathcal{P}_{k}(\tau)e^{-\alpha\tau}.
\]
\end{lemma}
\textit{Proof}. Let us make the inductive assumption that 
\[
H_{i}\leq C_{i},\ \ \
E_{i}\leq \mathcal{P}_{i}(\tau)e^{-\alpha\tau}
\]
for $i=1,...,k$, where the $C_{i}$ are constants and the 
$\mathcal{P}_{i}$ are polynomials. By Lemma \ref{lemma:hob} and 
(\ref{eq:eodec}), we know this to be true for $k=1$.
Note that if we let 
\[
\mathcal{F}_{l}=\frac{1}{2}\is (\partial_{\theta}^{l+1}P)^{2}d\theta,
\]
we have
\[
\frac{d\mathcal{F}_{l}}{d\tau}=\is \partial_{\theta}^{l+1}P
\partial_{\theta}^{l+1}\partial_{\tau}Pd\theta
\leq 2\mathcal{F}_{l}^{1/2}\mathcal{H}_{l+1}^{1/2},
\]
whence
\[
\mathcal{F}_{l}\leq C_{l}(1+\tau^{2})
\]
for $l=0,...,k-1$. Consequently, we have, using the definition
of $H_{k}$,
\begin{equation}\label{eq:psupo}
 \| \partial_{\theta}^{l}P\|_{C^{0}(S^{1},\mathbb{R})}\leq
C_{l}(1+\tau^{2})^{1/2},\ \ \
\| \partial_{\theta}^{k}P\|_{C^{0}(S^{1},\mathbb{R})}\leq
C_{k}e^{\a\tau/2}H_{k+1}^{1/2},
\end{equation}
\begin{equation}\label{eq:psupt}
 \| \partial_{\theta}^{k}P\|_{L^{2}(S^{1},\mathbb{R})}\leq
C_{k}(1+\tau^{2})^{1/2},\ \ \
\| \partial_{\theta}^{k+1}P\|_{L^{2}(S^{1},\mathbb{R})}\leq
C_{k}e^{\a\tau/2}H_{k+1}^{1/2},
\end{equation}
where $l=1,...,k-1$. Note that if $l\geq 1$, $\partial_{\theta}^{l}
\partial_{\tau}Q$ has to
have a zero on the circle since its average over the circle is zero.
Thus
\begin{equation}\label{eq:qse}
\|e^{P}\partial_{\theta}^{l}\partial_{\tau}Q\|_{C^{0}(S^{1},\mathbb{R})}\leq
\is |P_{\theta}e^{P}\partial_{\theta}^{l}
\partial_{\tau}Q+e^{P}\partial_{\theta}^{l+1}
\partial_{\tau}Q|d\theta
\end{equation}
\[
\leq \mathcal{Q}_{l}e^{-\a\tau/2}
+2\pi^{1/2}E_{l+1}^{1/2},
\]
for $l=1,...,k$, where $\mathcal{Q}_{l}$ is a polynomial in $\tau$ and
we have used the induction hypothesis and the first inequality in
(\ref{eq:psupt}) with $k=1$, which holds due to Lemma
\ref{lemma:hob}. If $l\leq k-1$, we thus get exponential decay of the
left hand side, due to the induction hypothesis. A similar estimate
holds if we replace $Q_{\tau}$ with $e^{-\tau}Q_{\theta}$. 

The idea of proof is the same as in Lemma \ref{lemma:hob}.
Consider 
\[
\frac{d\mathcal{E}_{k+1}}{d\tau}\leq
\is
\partial_{\theta}^{k+1}[e^{2P}(Q_{\tau}^{2}-e^{-2\tau}Q_{\theta}^{2})]
\partial_{\theta}^{k+1}\partial_{\tau}Pd\theta.
\]
By H\"{o}lder's inequality, we need to estimate expressions of the 
form 
\begin{equation}\label{eq:form}
\| \partial_{\theta}^{l_{1}}P\cdots
\partial_{\theta}^{l_{o}}P
e^{2P}\partial_{\theta}^{m}\partial_{\tau}Q
\partial_{\theta}^{n}\partial_{\tau}Q
\|_{L^{2}(S^{1},\mathbb{R})}
\end{equation}
where $l_{1}+...+l_{o}+m+n=k+1$, and the analogous expressions with
$Q_{\tau}$ replaced by $e^{-\tau}Q_{\theta}$. We wish to estimate this
by $H_{k+1}^{1/2}$ times something that decays exponentially. If the
biggest $l_{i}$ is $k+1$, we can use 
the second inequality in (\ref{eq:psupt}) and Lemma \ref{lemma:qdecay}.
If there are two $l_{i}$ which equal $k$, then $k=1$ and we can use 
Lemma \ref{lemma:qdecay}, the second inequality in (\ref{eq:psupo})
and the first inequality in (\ref{eq:psupt}) to obtain the desired
estimate. If there is only one $l_{i}$ equal to $k$ and the rest are
smaller, we can take out all factors $\partial_{\theta}^{l_{j}}P$
using the first or the second of (\ref{eq:psupo}). The remaining terms
are then estimated using Lemma \ref{lemma:qdecay} or (\ref{eq:eodec}).
If all the $l_{i}$ are less than or equal to $k-1$, we can
take out the derivatives of $P$ in the sup norm with a
polynomial bound in $\tau$, cf. the first inequality in
(\ref{eq:psupo}). Thus, we need not concern ourselves with derivatives
hitting $P$ in what follows. If one of $m$ and $n$ are less than or 
equal to $k-1$, we can take out the corresponding factor in the sup
norm, using (\ref{eq:qse}) and the fact that $E_{l+1}$ decays
exponentially for $l\leq k-1$. The problem which remains is the case
where $m=n=k$. Then $k=1$ and we can use (\ref{eq:eodec}) and 
(\ref{eq:qse}) to take out one factor in the sup norm and estimate the 
remaining $L^{2}$ norm by $E_{1}^{1/2}$. The argument for expressions
of the form (\ref{eq:form}) where one has replaced $Q_{\tau}$ with 
$e^{-\tau}Q_{\theta}$ is similar. 

Consider (\ref{eq:dekdt})-(\ref{eq:fkre}). Since $f_{0}=0$, we
inductively get the conclusion that $f_{k+1}$ is a sum of 
terms of the form 
\begin{equation}\label{eq:form2}
\partial_{\theta}^{m+1}\partial_{\tau}Pe^{2P}\partial_{\theta}^{l}
\partial_{\tau}Q,
\end{equation}
where $m+l=k$ and terms where one replaces $P_{\tau}$ with
$e^{-\tau}P_{\theta}$ and $Q_{\tau}$ with $e^{-\tau}Q_{\theta}$. Note
that the effect of the operator $\partial_{\theta}-2P_{\theta}$ on an
expression of the form $e^{2P}g$ is to only differentiate $g$. This is
the reason for (\ref{eq:form2}). Discarding the first term in 
(\ref{eq:dekdt}), we see that the estimate we wish to 
achieve is 
\[
\|e^{-P}f_{k+1}\|_{L^{2}(S^{1},\mathbb{R})}\leq Ce^{-\beta\tau}
H_{k+1}^{1/2}.
\]
Due to the form (\ref{eq:form2}), the relevant things to estimate
are thus
\begin{equation}\label{eq:intform}
\| \partial_{\theta}^{m+1}\partial_{\tau}Pe^{P}\partial_{\theta}^{l}
\partial_{\tau}Q\|_{L^{2}(S^{1},\mathbb{R})},
\end{equation}
where $l+m=k$. If $l\leq k-1$, we use (\ref{eq:qse}) to estimate 
$e^{P}\partial_{\theta}^{l}\partial_{\tau}Q$ in the sup norm. If
$l=k$, we take out $P_{\tau\theta}$ in the sup norm and use 
the induction hypothesis to conclude that what remains decays
exponentially. If $k=1$, we can estimate $P_{\tau\theta}$ by
$H_{k+1}^{1/2}$. Otherwise, it is bounded. The arguments 
for the terms where one replaces $P_{\tau}$ with
$e^{-\tau}P_{\theta}$ and $Q_{\tau}$ with $e^{-\tau}Q_{\theta}$
are similar.

Finally
\[
\frac{d }{d\tau}[\frac{1}{2}e^{-\a\tau}\is (\partial_{\theta}^{k+1}P)^{2}
d\theta]
\leq e^{-\a\tau}\is \partial_{\theta}^{k+1}P
\partial_{\theta}^{k+1}\partial_{\tau}Pd\theta
\leq 2e^{-\a\tau/2}H_{k+1}.
\]
In consequence, we have the desired inequality for $H_{k+1}'$ and 
we conclude that $H_{k+1}$ is bounded. In order to complete the 
induction, we need to prove the decay of $E_{k+1}$. Due to
(\ref{eq:dekdt}), we only need to prove that expressions of the 
form (\ref{eq:intform}) can be estimated by a polynomial times
$e^{-\a\tau/2}$ (and the analogous expressions where $\partial_{\tau}$
is replaced with $e^{-\tau}\partial_{\theta}$). If $l\leq k-1$, we can
use (\ref{eq:qse}) and
the boundedness of $H_{k+1}$. Since $P_{\tau\theta}$ is bounded
in the sup norm due to the boundedness of $H_{2}$, and since 
$E_{k}$ has the desired decay, we can also deal with the case
$l=k$. The lemma follows. $\hfill\Box$

\textit{Proof of Theorem \ref{thm:expansions}}. 
The conditions of Corollary \ref{cor:continuation}
are satisfied. By Lemma \ref{lemma:hkb}, we conclude
that $\mathcal{E}_{k}$ is bounded and that $E_{k}$ decays as 
$e^{-\a\tau}$ times a polynomial for all $k$. Furthermore, for 
all $k$, there is a polynomial $\mathcal{Q}_{k}$ in $\tau$
such that for all $k$
\[
\|e^{P}\partial_{\theta}^{k}\partial_{\tau}Q\|_{C^{0}(S^{1},
\mathbb{R})}+\|e^{P-\tau}\partial_{\theta}^{k+1}Q\|_{C^{0}(S^{1},
\mathbb{R})}\leq \mathcal{Q}_{k}(\tau)e^{-\a\tau/2}.
\]
This follows from (\ref{eq:qse}) and Lemma \ref{lemma:hkb}.
When one has these estimates, it is not 
so difficult to combine them with the equations in order to 
get the conclusions of the theorem. However, the arguments
require almost two pages, so we refer the reader to pp. 23-25
of \cite{jagg} for the details. $\hfill\Box$

\section{The disc model}\label{section:dm}

We begin by proving a result analogous to Lemma \ref{lemma:decay}
in the disc model. 

\begin{lemma}\label{lemma:rhobo}
Let $z$ be a solution to (\ref{eq:edisc}) and assume that there is 
a $T$ such that $\rho(\tau,\theta)\leq \tau-2$ for $(\tau,\theta)
\in [T,\infty)\times S^{1}$, where $\rho$ is defined in
(\ref{eq:rhodef}). Then there is a $v\in C^{0}(S^{1},\mathbb{R}^{2})$
with $|v|\leq 1$ and a $C$ such that for $\tau\geq T$, 
\[
\|\frac{1}{\tau}\frac{z(\tau,\cdot)}{|z(\tau,\cdot)|}
\rho(\tau,\cdot)-v\|_{C^{0}(S^{1},\mathbb{R}^{2})}+
\|\frac{2z_{\tau}(\tau,\cdot)}{1-|z(\tau,\cdot)|^{2}}-
v\|_{C^{0}(S^{1},\mathbb{R}^{2})}
\]
\[
+e^{-\tau}\|\frac{2z_{\theta}(\tau,\cdot)}{1-|z(\tau,\cdot)|^{2}}
\|_{C^{0}(S^{1},\mathbb{R}^{2})}\leq C\tau^{-1}.
\]
\end{lemma}
\textit{Proof}. If $z$ is a solution of (\ref{eq:edisc}), let 
\begin{equation}\label{eq:aotdef}
\mathcal{A}_{1}=\frac{1}{2}e^{\tau}
|z_{\tau}+e^{-\tau}z_{\theta}|_{D}^{2},
\ \ \
\mathcal{A}_{2}=\frac{1}{2}e^{\tau}
|z_{\tau}-e^{-\tau}z_{\theta}|_{D}^{2},
\end{equation}
with notation as in (\ref{eq:dn}), and
\[
\mathcal{C}_{1}=-\frac{1}{2\tau}e^{\tau}(\partial_{\tau}+e^{-\tau}
\partial_{\theta})(\rho)^{2}+\frac{1}{2}e^{\tau}\frac{\rho^{2}}{\tau^{2}},
\ \ \
\mathcal{C}_{2}=-\frac{1}{2\tau}e^{\tau}(\partial_{\tau}-e^{-\tau}
\partial_{\theta})(\rho)^{2}+\frac{1}{2}e^{\tau}\frac{\rho^{2}}{\tau^{2}}.
\]
Note that $\mathcal{C}_{1}$ and $\mathcal{C}_{2}$ are smooth functions.
If we let 
\[
f=\frac{1}{2\tau}(1-|z|^{2})\frac{z}{|z|}\rho,
\]
which is a smooth function, then
\begin{equation}\label{eq:bodef}
\mathcal{B}_{1}=\mathcal{A}_{1}+\mathcal{C}_{1}
=\frac{1}{2}e^{\tau}
|z_{\tau}-f+e^{-\tau}z_{\theta}|_{D}^{2}
\end{equation}
and similarly for $\mathcal{B}_{2}=\mathcal{A}_{2}+
\mathcal{C}_{2}$. Consequently $\mathcal{B}_{1}$ and $\mathcal{B}_{2}$
are non-negative. We have 
\begin{equation}\label{eq:aotpr}
(\partial_{\tau}-e^{-\tau}\partial_{\theta})\mathcal{A}_{1}=
(\partial_{\tau}+e^{-\tau}\partial_{\theta})\mathcal{A}_{2}=
\frac{1}{2}e^{\tau}
[|z_{\tau}|_{D}^{2}-e^{-2\tau}|z_{\theta}|_{D}^{2}].
\end{equation}
Let us compute 
\[
(\partial_{\tau}-e^{-\tau}\partial_{\theta})\mathcal{C}_{1}
=\frac{1}{\tau^{2}}e^{\tau}\partial_{\tau}(\rho^{2})-
\frac{1}{2\tau}e^{\tau}\partial_{\tau}(\rho^{2})+\frac{1}{2}
e^{\tau}\frac{\rho^{2}}{\tau^{2}}-\frac{1}{\tau}e^{\tau}
\frac{\rho^{2}}{\tau^{2}}
\]
\[
-\frac{1}{2\tau}e^{\tau}
(\partial_{\tau}^{2}-e^{-2\tau}\partial_{\theta}^{2})(\rho^{2}).
\]
If $|z|>0$, we can compute
\begin{equation}\label{eq:triv}
\frac{1}{2}(\partial_{\tau}^{2}-e^{-2\tau}\partial_{\theta}^{2})(\rho^{2})
=(\rho_{\tau}^{2}-e^{-2\tau}\rho_{\theta}^{2})+
\rho (\rho_{\tau\tau}-e^{-2\tau}\rho_{\theta\theta}),
\end{equation}
and that 
\[
\sinh\rho (\rho_{\tau\tau}-e^{-2\tau}\rho_{\theta\theta})=
\cosh\rho [|z_{\tau}|_{D}^{2}-e^{-2\tau}
|z_{\theta}|_{D}^{2}-\rho_{\tau}^{2}+e^{-2\tau}\rho_{\theta}^{2}].
\]
Observe that 
\begin{equation}\label{eq:div}
\rho_{\tau}^{2}+\sinh^{2}\rho |\partial_{\tau}(\frac{z}{|z|})|^{2}=
|z_{\tau}|_{D}^{2}
\end{equation}
and similarly for the $\theta$ derivatives. Consequently
\[
|z_{\tau}|_{D}^{2}-\rho_{\tau}^{2}\geq 0.
\]
Since $\rho\cosh\rho\geq \sinh\rho$, we conclude that 
\[
 -\rho (\rho_{\tau\tau}-e^{-2\tau}\rho_{\theta\theta})\leq
-|z_{\tau}|_{D}^{2}
+\rho_{\tau}^{2}-e^{-2\tau}\rho_{\theta}^{2}
+\frac{\rho\cosh\rho}{\sinh\rho} 
e^{-2\tau}|z_{\theta}|_{D}^{2}
\]
Combining this with (\ref{eq:triv}), we get
\[
-\frac{1}{2}
(\partial_{\tau}^{2}-e^{-2\tau}\partial_{\theta}^{2})(\rho^{2})
\leq 
-|z_{\tau}|_{D}^{2}
+\frac{\rho\cosh\rho}{\sinh\rho} 
e^{-2\tau}|z_{\theta}|_{D}^{2}.
\]
Since $\rho\cosh\rho/\sinh\rho\leq\tau-1$ if $\rho\leq\tau-2$,
we get 
\[
-\frac{1}{2\tau}e^{\tau}
(\partial_{\tau}^{2}-e^{-2\tau}\partial_{\theta}^{2})(\rho^{2})
\leq
e^{-\tau}|z_{\theta}|_{D}^{2}
-\frac{1}{\tau}e^{\tau}
[|z_{\tau}|_{D}^{2}+e^{-2\tau}|z_{\theta}|_{D}^{2}].
\]
We conclude that if $\rho\leq\tau-2$, then
\begin{equation}\label{eq:bopr}
 (\partial_{\tau}-e^{-\tau}\partial_{\theta})\mathcal{B}_{1}\leq
(\frac{1}{2}-\frac{1}{\tau})e^{\tau}
[|z_{\tau}|_{D}^{2}+e^{-2\tau}|z_{\theta}|_{D}^{2}]
+\frac{1}{\tau^{2}}e^{\tau}\partial_{\tau}(\rho^{2})
\end{equation}
\[
-\frac{1}{2\tau}e^{\tau}\partial_{\tau}(\rho^{2})
+\frac{1}{2}
e^{\tau}\frac{\rho^{2}}{\tau^{2}}-\frac{1}{\tau}e^{\tau}
\frac{\rho^{2}}{\tau^{2}}=(\frac{1}{2}-\frac{1}{\tau})
(\mathcal{B}_{1}+\mathcal{B}_{2}).
\]
Similarly,
\begin{equation}\label{eq:btpr}
(\partial_{\tau}+e^{-\tau}\partial_{\theta})\mathcal{B}_{2}\leq
(\frac{1}{2}-\frac{1}{\tau})(\mathcal{B}_{1}+\mathcal{B}_{2}).
\end{equation}
The above derivation was carried out under the assumption that 
$|z|>0$. However, if $z(\tau,\theta)=0$, then $\rho^{2}$ can 
be replaced with $4|z|^{2}$ when calculating second derivatives.
Thus, at the point $(\tau,\theta)$, we have 
\[
(\partial_{\tau}^{2}-e^{-2\tau}\partial_{\theta}^{2})(\rho^{2})
=8(|z_{\tau}|^{2}-e^{-2\tau}|z_{\theta}|^{2}).
\]
Using this observation, one can see that (\ref{eq:bopr}) and 
(\ref{eq:btpr}) hold regardless of whether $|z|=0$ or not. 
Let 
\[
G(\tau)=e^{-\tau}\sup_{\theta}\mathcal{B}_{1}(\tau,\theta)+
e^{-\tau}\sup_{\theta}\mathcal{B}_{2}(\tau,\theta),
\]
and assume that there is a $T$ such that $\rho(\tau,\theta)\leq\tau-2$
for all $(\tau,\theta)\in [T,\infty)\times S^{1}$. Then an argument
similar to the proof of Lemma \ref{lemma:decay} shows that 
\[
G(\tau)\leq G(\tau_{0})\left(\frac{\tau_{0}}{\tau}\right)^{2}
\]
if $\tau\geq\tau_{0}\geq T$. Assuming $|z|>0$, we can use 
(\ref{eq:div}) in order to obtain
\[
e^{-\tau}\mathcal{B}_{1}+e^{-\tau}\mathcal{B}_{2}=
(\rho_{\tau}-\frac{1}{\tau}\rho)^{2}+\sinh^{2}\rho 
|\partial_{\tau}(\frac{z}{|z|})|^{2}+
e^{-2\tau}|z_{\theta}|_{D}^{2}.
\]
Consequently, there is a constant $C$ such that for $\tau\geq T$
\[
(\rho_{\tau}-\frac{1}{\tau}\rho)^{2}+\sinh^{2}\rho 
|\partial_{\tau}(\frac{z}{|z|})|^{2}\leq C\tau^{-2}
\]
assuming $|z|>0$. Define $g$ by
\[
g=\frac{1}{\tau}\frac{z}{|z|}\rho.
\]
Let us compute, assuming $|z|>0$, 
\[
\partial_{\tau}g=\frac{z}{|z|}\frac{1}{\tau}(\rho_{\tau}-\frac{1}{\tau}\rho)
+\frac{1}{\tau}\partial_{\tau}(\frac{z}{|z|})\rho.
\]
Since $\rho/\sinh\rho$ is bounded, we get the conclusion that 
\begin{equation}\label{eq:gbound}
|\partial_{\tau}g|\leq C\tau^{-2}
\end{equation}
for $\tau\geq T$, assuming $|z|>0$. If $z=0$, we get 
\[
\partial_{\tau}g=\frac{2}{\tau}z_{\tau}.
\]
However, in this case 
\[
e^{-\tau}\mathcal{B}_{1}+e^{-\tau}\mathcal{B}_{2}=
4|z_{\tau}|^{2}+4e^{-2\tau}|z_{\theta}|^{2}.
\]
Consequently (\ref{eq:gbound}) holds regardless of whether
$z$ is zero or not. If $\tau_{2}\geq\tau_{1}\geq T$, we thus get
the conclusion that 
\[
\| g(\tau_{2},\cdot)-g(\tau_{1},\cdot)\|_{C^{0}(S^{1},\mathbb{R})}
\leq C\tau_{1}^{-1}.
\]
In other words, there is a $v\in C^{0}(S^{1},\mathbb{R}^{2})$
such that 
\[
\|g(\tau,\cdot)-v\|_{C^{0}(S^{1},\mathbb{R}^{2})}\leq C\tau^{-1}.
\]
Note that $|v|\leq 1$, since $|g|\leq 1$ for $\tau\geq T$. Since
\[
e^{-\tau}\mathcal{B}_{1}+e^{-\tau}\mathcal{B}_{2}=
|\frac{2z_{\tau}}{1-|z|^{2}}-g|^{2}+e^{-2\tau}|z_{\theta}|_{D}^{2},
\]
the lemma follows. $\hfill\Box$

\textit{Proof of Theorem \ref{thm:td}}. 
Consider $\mathcal{A}_{1}$ and $\mathcal{A}_{2}$
defined in (\ref{eq:aotdef}). Due to (\ref{eq:aotpr}), we have 
\[
(\partial_{\tau}-e^{-\tau}\partial_{\theta})\mathcal{A}_{1}=
(\partial_{\tau}+e^{-\tau}\partial_{\theta})\mathcal{A}_{2}\leq
\frac{1}{2}(\mathcal{A}_{1}+\mathcal{A}_{2}).
\]
This can be used to conclude that the object appearing on the left
hand side of (\ref{eq:object}) is monotonically decaying to the 
future. The argument is similar to the proof of Lemma
\ref{lemma:decay}. Consequently, $|z_{\tau}|_{D}\leq 1-\a$ to the 
future. Due to (\ref{eq:div}), this means that 
$\rho(\tau,\theta)\leq (1-\a)\tau+C$ for $\tau\geq \tau_{0}$.
At late enough times, the conditions of Lemma \ref{lemma:rhobo}
are thus fulfilled. The theorem follows. $\hfill\Box$

\section{Local results}

Here we wish to state some results that are local in the spatial
coordinate. We also wish to relate the pictures in the different 
models of hyperbolic space. Note that we have the three models
of hyperbolic space (\ref{eq:grdef}), (\ref{eq:ghdef}) and 
(\ref{eq:gddef}) which are related by the isometries 
(\ref{eq:phirdef}) and (\ref{eq:phihdef}).

\begin{lemma}\label{lemma:loco}
Consider a solution to (\ref{eq:edisc}). Let $\theta_{0}\in S^{1}$
and assume that there are 
$\gamma,\epsilon>0$, $T$ and $v\in
C^{0}(I_{\epsilon},\mathbb{R}^{2})$, where $I_{\epsilon}
=(\theta_{0}-\epsilon,\theta_{0}+\epsilon)$, 
$2\gamma\leq |v|\leq 1-2\gamma $ in $I_{\epsilon}$ and 
\begin{equation}\label{eq:decno}
\|\frac{1}{\tau}\frac{z(\tau,\cdot)}{|z(\tau,\cdot)|}
\rho(\tau,\cdot)-v\|_{C^{0}(I_{\epsilon},\mathbb{R}^{2})}
\leq C\tau^{-1}
\end{equation}
for $\tau\geq T$. Then there is an $\eta>0$ and an
isometry $\phi$ from $(D,g_{D})$ to $(\mathbb{R}^{2},g_{R})$
such that if $(Q,P)=\phi\circ z$, then the conclusions of 
Theorem \ref{thm:expansions} hold if we replace $S^{1}$ with
$I_{\eta}$.
\end{lemma}
\textit{Proof}. Using the isometries (\ref{eq:phirdef}) and 
(\ref{eq:phihdef}), we conclude that the 
relation between variables in the disc $z$ and the $PQ$-variables
given by
\begin{equation}\label{eq:QP}
(Q,P)=[-\frac{2\mathrm{Im}z}{1+|z|^{2}-2\mathrm{Re}z},
-\ln (1-|z|^{2})+\ln (1+|z|^{2}-2\mathrm{Re}z)]
\end{equation}
is an isometry. Using (\ref{eq:rhodef}) and (\ref{eq:QP}), we have 
\begin{equation}\label{eq:pex}
P=\rho-2\ln (1+|z|)+\ln (1+|z|^{2}-2\mathrm{Re}z).
\end{equation}
By the assumptions of the lemma, there is a positive $\gamma$ such that 
\begin{equation}\label{eq:vb}
2\gamma\leq |v(\theta)|\leq 1-2\gamma
\end{equation}
for $\theta\in I_{\epsilon}$
and by making $\epsilon$ smaller, if necessary, we can assume that the
image of $I_{\epsilon}$ under $v/|v|$ is contained
in an angle interval of length less that $\pi$. We can also assume that 
$\rho\geq 1$ in $[\tau_{0},\infty)\times I_{\epsilon}$. By
(\ref{eq:decno}) we have 
\[
\| \frac{z(\tau,\cdot)}{|z(\tau,\cdot)|}-
\frac{v}{|v|}\|_{C^{0}(I_{\epsilon},\mathbb{R}^{2})}\leq C\tau^{-1}
\]
for $\tau\geq \tau_{0}$. Consequently, we can assume that $z/|z|$
is contained in an angle interval of $3\pi/2$. Thus we can perform
a rotation to the solution in order to ensure that 
\begin{equation}\label{eq:zazb}
|z/|z|-1|\geq c 
\end{equation}
for some positive constant $c$ and $(\tau,\theta)\in [\tau_{0},
\infty)\times I_{\epsilon}$. From now on, we will assume that such
a rotation has been carried out. By (\ref{eq:pex}) and
(\ref{eq:zazb}), we conclude that $P=\rho+O(1)$. Combining this
with (\ref{eq:decno}) and (\ref{eq:vb}), we get the conclusion that
there is a $T$ such that for $(\tau,\theta)\in [T,\infty)\times 
I_{\epsilon}$,
\[
1\leq P(\tau,\theta)\leq\tau-1,\ \ \
\gamma\leq\frac{P(\tau,\theta)}{\tau}\leq 1-\gamma.
\]
We can now carry out a localized version of the argument presented
in the proof of Lemma \ref{lemma:decay}. Instead of considering
the set $[T,\infty)\times S^{1}$ we consider $(\tau,\theta)$ such
that $\tau\geq T$ and 
\[
\theta\in I_{\epsilon,T,\tau}=(\theta_{0}-\epsilon-e^{-\tau}+e^{-T},
\theta_{0}+\epsilon+e^{-\tau}-e^{-T}),
\]
where we assume that $T$ is big enough that $e^{-T}\leq\epsilon/2$.
We conclude that if 
\[
F_{\epsilon,T}(\tau)=
\frac{1}{2}e^{-\tau}
\sup_{\theta\in I_{\epsilon,T,\tau}}\mathcal{A}(\tau,\theta)+
\frac{1}{2}e^{-\tau}
\sup_{\theta\in I_{\epsilon,T,\tau}}\mathcal{B}(\tau,\theta),
\]
where $\mathcal{A}$ and $\mathcal{B}$ are defined in 
(\ref{eq:abdef}), then
\[
F_{\epsilon,T}(\tau)\leq C\tau^{-2}.
\]
For any $\xi>0$ there is a $T_{2}$ such that if $\tau\geq T_{2}$, 
then we can change the initial data outside 
of $(\theta_{0}-\epsilon/4,\theta_{0}+\epsilon/4)$ so that 
$F(\tau)\leq \xi$, where $F$ is defined in (\ref{eq:fdef})
and $\gamma\leq P(\tau,\cdot)/\tau\leq 1-\gamma$. Let us be 
explicit concerning the change of the initial data. Let 
$\chi\in C^{\infty}(S^{1},\mathbb{R})$ be such that $\chi(\theta)
=1$ for $\theta$ outside of $(\theta_{0}-\epsilon/2,
\theta_{0}+\epsilon/2)$ and $\chi(\theta)
=0$ for $\theta\in (\theta_{0}-\epsilon/4,
\theta_{0}+\epsilon/4)$. Since $Q$ is bounded in the relevant
set, due to (\ref{eq:QP}) and (\ref{eq:zazb}), the modifications 
\[
 \tilde{P}=(1-\chi)P+\chi\gamma\tau,\ \
\tilde{P}_{\tau}=(1-\chi)P_{\tau}+\chi\gamma,\ \
\tilde{Q}=(1-\chi)Q,\ \ \tilde{Q}_{\tau}=(1-\chi)Q_{\tau}\ \
\]
yield the desired conclusion for large enough $\tau$. 
By modifying the initial data as above at a late enough time $T_{3}$, 
the conditions of Theorem \ref{thm:expansions} will be 
fulfilled. Furthermore, we can assume that we have the original 
solution in the strip $[T_{3},\infty)\times (\theta_{0}-\epsilon/8,
\theta_{0}+\epsilon/8)$. Consequently, we get the conclusions of 
Theorem \ref{thm:expansions} localized to the above mentioned strip.
$\hfill\Box$

\begin{lemma}\label{lemma:rholbo}
Consider a solution to (\ref{eq:edisc}).
Let $\theta_{0}\in S^{1}$ be a fixed angle, and assume that there is 
a $T$ and a $0<\gamma<1$ such that 
\[
\frac{\rho(\tau,\theta_{0})}{\tau}\leq 1-\gamma
\]
for all $\tau\geq T$. Then there is an $\eta>0$ and a 
$v\in C^{0}(I_{\eta},\mathbb{R}^{2})$ such that for $\tau\geq T$,
\[
\|\frac{1}{\tau}\frac{z(\tau,\cdot)}{|z(\tau,\cdot)|}
\rho(\tau,\cdot)-v\|_{C^{0}(I_{\eta},\mathbb{R}^{2})}+
\|\frac{2z_{\tau}(\tau,\cdot)}{1-|z(\tau,\cdot)|^{2}}-
v\|_{C^{0}(I_{\eta},\mathbb{R}^{2})}
\]
\[
+e^{-\tau}\|\frac{2z_{\theta}(\tau,\cdot)}{1-|z(\tau,\cdot)|^{2}}
\|_{C^{0}(I_{\eta},\mathbb{R}^{2})}\leq C\tau^{-1}.
\]
\end{lemma}
\textit{Proof}. Let 
\[
\mathcal{S}_{\tau}=(\theta_{0}-2e^{-\tau},\theta_{0}+2e^{-\tau}),
\]
where we assume that $\tau$ is big enough that $2e^{-\tau}\leq\pi/2$.
We know that 
\[
|z_{\tau}|_{D}^{2}+e^{-2\tau}|z_{\theta}|_{D}^{2}\leq C
\]
for $(\tau,\theta)\in [T,\infty)\times S^{1}$ and that (\ref{eq:div})
holds, and a similar division for the spatial derivatives, assuming
$|z|>0$. Consequently, 
\[
|\frac{\rho(\tau,\theta_{1})}{\tau}-\frac{\rho(\tau,\theta_{2})}{\tau}|
\leq \frac{C}{\tau},
\]
assuming $\theta_{i}\in \mathcal{S}_{\tau}$. Thus there is a 
$T_{1}\geq T$ such that (assuming $3\beta=\gamma$)
\begin{equation}\label{eq:init}
\rho(\tau,\theta)/\tau\leq 1-2\beta\ \ \ \mathrm{and}\ \ \
\rho(\tau,\theta)\leq \tau-2
\end{equation}
for $\tau\in [T_{1},\infty)$ and $\theta\in\mathcal{S}_{\tau}$.
The idea of the argument is as follows. First we consider 
a quantity similar to the $G$ introduced in the proof of Lemma 
\ref{lemma:rhobo}, the only difference being that we take 
supremum over $\mathcal{S}_{\tau}$ instead of $S^{1}$. At a 
late enough time, say $\tau_{1}$, the quantity analogous to $G$ has
become small enough that we can close the argument and make statements
concerning the domain determined by $\mathcal{S}_{\tau_{1}}$.
This domain contains an open subset of the singularity.

Let $\mathcal{B}_{1}$ be defined as in (\ref{eq:bodef}) and similarly
for $\mathcal{B}_{2}$. Let $\tau\geq \tau_{0}\geq T_{1}$.
Define
\[
 L_{1}(u,\theta)=\mathcal{B}_{1}(u,\theta+e^{-u}-e^{-\tau})
\ \ \ \mathrm{and}\ \ \
L_{2}(u,\theta)=\mathcal{B}_{2}(u,\theta-e^{-u}+e^{-\tau}),
\]
where $\theta\in\sm_{\tau}$. Note that 
\[
\theta+e^{-u}-e^{-\tau}, \theta-e^{-u}+e^{-\tau}
\in \sm_{u}
\]
for $u\leq \tau$ and $\theta\in \sm_{\tau}$. Let
\[
\hat{L}_{i}(u)=\sup_{\theta\in\sm_{u}}\mathcal{B}_{i}(u,\theta)\ \ \
\mathrm{and}\ \ \
\hat{L}=\hat{L}_{1}+\hat{L}_{2}.
\]
Due to (\ref{eq:bopr}) we have, for $\theta\in\sm_{\tau}$,
\[
 \mathcal{B}_{1}(\tau,\theta)=L_{1}(\tau_{0},\theta)
+\int_{\tau_{0}}^{\tau}[(\partial_{u} -e^{-u}\partial_{\theta})
\mathcal{B}_{1}](u,\theta+e^{-u}-e^{-\tau})d u
\]
\[
\leq
\hat{L}_{1}(\tau_{0})+\int_{\tau_{0}}^{\tau}(\frac{1}{2}-
\frac{1}{u})(\hat{L}_{1}+\hat{L}_{2})d u.
\]
Taking the supremum over $\theta\in\sm_{\tau}$ and adding a similar
estimate for $\mathcal{B}_{2}$, we get the conclusion that 
\[
\hat{L}(\tau)\leq\hat{L}(\tau_{0})+
\int_{\tau_{0}}^{\tau}(1-\frac{2}{u})\hat{L}(u)d u.
\]
Consequently,
\[
e^{-\tau}\hat{L}(\tau)\leq e^{-\tau_{0}}\hat{L}(\tau_{0})
\frac{\tau_{0}^{2}}{\tau^{2}}.
\]
Let us consider a similar argument on a different set. Let 
\[
\sm_{\tau_{1},\tau}= (\theta_{0}-e^{-\tau}-e^{-\tau_{1}},
\theta_{0}+e^{-\tau}+e^{-\tau_{1}}).
\]
Note that $\sm_{\tau_{1},\tau_{1}}=\sm_{\tau_{1}}$. For
$\theta\in\sm_{\tau_{1},\tau}$, let 
\[
 K_{1}(u,\theta)=\mathcal{B}_{1}(u,\theta+e^{-u}-e^{-\tau})
\ \ \ \mathrm{and}\ \ \
K_{2}(u,\theta)=\mathcal{B}_{2}(u,\theta-e^{-u}+e^{-\tau}).
\]
For $u\leq\tau$, we have 
\[
\theta+e^{-u}-e^{-\tau},\theta-e^{-u}+e^{-\tau}\in
\sm_{\tau_{1},u}.
\]
Letting 
\[
\hat{K}_{i}(u)=\sup_{\theta\in\sm_{\tau_{1},u}}
\mathcal{B}_{i}(u,\theta)\ \ \ \mathrm{and}\ \ \
\hat{K}=\hat{K}_{1}+\hat{K}_{2},
\]
we can argue similarly to the above in order to obtain
\begin{equation}\label{eq:ghdecay}
e^{-\tau}\hat{K}(\tau)\leq e^{-\tau_{1}}\hat{K}(\tau_{1})
\frac{\tau_{1}^{2}}{\tau^{2}},
\end{equation}
assuming $\rho(u,\theta)\leq u-2$ for $u\in [\tau_{1},\tau]$
and $\theta\in\sm_{\tau_{1},u}$. Note that $\hat{K}(\tau_{1})
=\hat{F}(\tau_{1})$. The problem is of course that we cannot
assume that $\rho\leq \tau-2$ in the relevant set. Let us
assume that $\tau_{1}$ is big enough so that 
\begin{equation}\label{eq:init2}
e^{-\tau_{1}}
\hat{K}(\tau_{1})\leq\beta^{2},\ \ \ 
\tau_{1}\beta\geq 2
\end{equation}
and $\tau_{1}\geq T_{1}$. Due to (\ref{eq:init}), we know that 
$\rho(\tau_{1},\theta)/\tau_{1}\leq 1-2\beta$ for $\theta\in
\mathcal{S}_{\tau_{1}}$. Let
\[
\sm=\{ \tau\in [\tau_{1},\infty): s\in [\tau_{1},\tau], 
\theta\in\sm_{\tau_{1},s} \Rightarrow \frac{\rho(s,\theta)}{s}\leq
1-\beta\}.
\]
Note that $\sm$ is closed, connected and non-empty. Let us prove that 
it is open. Let $\tau\in\sm$. Then (\ref{eq:ghdecay}) is applicable 
in $[\tau_{1},\tau]$ due to (\ref{eq:init2}). Consequently, if
$\theta\in\sm_{\tau_{1},\tau}$, then
\[
|\frac{\rho(\tau,\theta)}{\tau}-\frac{\rho(\tau_{1},\theta)}{\tau_{1}}|
=|\int_{\tau_{1}}^{\tau}\frac{1}{s}(\rho_{s}-
\frac{\rho}{s})d s| \leq \int_{\tau_{1}}^{\tau}\beta
\frac{\tau_{1}}{s^{2}}d s=\beta(1-\frac{\tau_{1}}{\tau}).
\]
This implies, using (\ref{eq:init}), that 
\[
\frac{\rho(\tau,\theta)}{\tau}\leq 1-\beta(1+\frac{\tau_{1}}{\tau}).
\]
Consequently, $\sm$ is open. The conclusions of the lemma follow
by an argument similar to the proof of Lemma \ref{lemma:rhobo}.
$\hfill\Box$

\section*{Acknowledgements}

This work was partly carried out when the author was enjoying the 
the hospitality of Rutgers. The author would like to express his
gratitude to A. Shadi Tahvildar-Zadeh for the invitation and many
stimulating discussions.


\begin{thebibliography}{1}
\bibitem{bag} Berger B and Garfinkle D 1998 Phenomenology of the Gowdy
	universe on $T^{3}\times \mathbb{R}$ {\em Phys. Rev. D}
	{\bf 57} 1767--77
\bibitem{chr1} Chru\'{s}ciel P T 1990 On spacetimes with $U(1)\times
U(1)$ symmetric compact Cauchy surfaces {\em Ann. Phys. NY}
{\bf 202} 100--50
\bibitem{chr2} Chru\'{s}ciel P T 1991 On uniqueness in the large of 
solutions of Einstein's equations ('strong cosmic censorship')
{\em Proc. Centre for Mathematical Analysis} vol 27 Australian
National University
\bibitem{gowdy} Gowdy R H 1974 Vacuum spacetimes with two-parameter
spacelike isometry groups and compact invariant hypersurfaces: 
Topologies and boundary conditions {\em Ann. Phys. NY} {\bf 83}
203--41
\bibitem{grub} Grubi\v{s}i\'{c} B and Moncrief V 1993 Asymptotic
behaviour of the $T^{3}\times \mathbb{R}$ Gowdy space-times
{\it Phys. Rev. D} {\bf 47} 2371--82
\bibitem{jam} Isenberg J and Moncrief V 1990 Asymptotic behaviour
	of the gravitational field and the nature of singularities
	in Gowdy space times {\em Ann. Phys} {\bf 199} 84--122
\bibitem{mon} Moncrief V 1981 Global properties of Gowdy spacetimes
with $T^{3}\times \mathbb{R}$ topology {\em Ann. Phys. NY} {\bf 132}
87--107
\bibitem{kar1} Kichenassamy S and Rendall A 1998 Analytic
description of singularities in Gowdy spacetimes {\it Class.
Quantum Grav.} {\bf 15} 1339--55
\bibitem{r} Rendall A 2000 Fuchsian analysis of singularities 
in Gowdy spacetimes beyond analyticity {\it Class. Quantum Grav.}
{\bf 17} 3305--16
\bibitem{raw} Rendall A and Weaver M 2001 Manufacture of Gowdy
        spacetimes with spikes {\em Class. Quantum Grav.} {\bf 18}
	2959--76
\bibitem{jagg} Ringstr\"{o}m H 2002 On Gowdy vacuum spacetimes
{\it Preprint} gr-qc/0204044 To appear in the Mathematical proceedings
of the Cambridge Philosophical Society 
\end{thebibliography}
\end{document}